\documentclass[reprint, amsmath, amssymb, aps, prb, floatfix]{revtex4-1}

\usepackage{graphicx}
\usepackage{hyperref}
\hypersetup{colorlinks = true, citecolor = blue, breaklinks = true}
\usepackage[version=4]{mhchem}
\usepackage[resetlabels]{multibib}
\newcites{SI}{References}

\begin{document}

\title{Multiferroicity and 180$^\circ$ domain switching in \ce{LaFeO3} via Antisite Defects}

\author{Souren Majani}
\affiliation{Department of Chemistry and Physics of Materials, University of Salzburg, Jakob-Haringer-Strasse 2a, 5020 Salzburg, Austria}

\author{Harish K. Singh}
\affiliation{Department of Chemistry and Physics of Materials, University of Salzburg, Jakob-Haringer-Strasse 2a, 5020 Salzburg, Austria}

\author{Ulrich Aschauer}
\email{ulrich.aschauer@plus.ac.at}
\affiliation{Department of Chemistry and Physics of Materials, University of Salzburg, Jakob-Haringer-Strasse 2a, 5020 Salzburg, Austria}

\date{\today}

\begin{abstract}
Materials with coexisting and coupled ferroelectric and magnetic orders are rare. Here we show, using density functional theory calculations, that inducing \ce{Fe_{La}} antisites into non-ferroelectric and antiferromagnetic \ce{LaFeO3} renders the material at the same time ferroelectric and ferrimagnetic. Even more excitingly, we observe a direct coupling between the ferroelectric and ferrimagnetic polarization, the latter being switchable by the former. While on average the magnetic moments of antisites would cancel, we envision that preparing defective \ce{LaFeO3} under simultaneous electric and magnetic fields will lead to a net magnetic moment due to magnetic domain reconfiguration. Moreover, ferroelectric switching under a static magnetic field can lead to 180$^\circ$ switching of the antiferromagnetic order in \ce{LaFeO3}.
\end{abstract}

\maketitle

%%%%%%%%%%%%%%%%%%%%%%%%%%%%%%%%%%%%%%%%%%%%%%%%%%%%%%%%%%%%%%%%%%%%%%%%%%%%%%%%%
\section{Introduction}
%%%%%%%%%%%%%%%%%%%%%%%%%%%%%%%%%%%%%%%%%%%%%%%%%%%%%%%%%%%%%%%%%%%%%%%%%%%%%%%%%

Multiferroic materials exhibiting ferroelectric and ferromagnetic properties are at the forefront of next-generation memory technology. Their unique ability to couple electric and magnetic orders opens avenues for innovative applications, such as four-state memory elements that encode data using both electric and magnetic bits~\cite{son2013four, ruan2016four}. Moreover, the coupling between ferromagnetic and ferroelectric orders in multiferroic materials enables the control of magnetic states using electric fields, eliminating the need for current-generated magnetic fields. This approach can significantly reduce the size of memory elements while accelerating data read and write processes. In addition, utilizing electric fields for magnetization control is more energy efficient than traditional methods, potentially reducing energy consumption by several orders of magnitude~\cite{ramesh2021electric}. Indeed, multiferroic systems have demonstrated all electric field controlled logic operations, paving the way for scalable, low power memory in logic applications~\cite{li2023energy}.

Perovskite oxides, with the general formula \ce{ABO3}, are notable for their ability to host a variety of elements at both the A and B site cation positions~\cite{pena2001chemical}. This compositional flexibility allows for fine-tuning their structural, electronic, and magnetic properties. Beyond compositional adjustments, the deliberate introduction of point defects such as vacancies, interstitials, and antisite substitutions serves as a strategy to modify and enhance the functional properties of perovskite oxides~\cite{Ricca2022}. Historically, defects in ferroelectric materials were considered detrimental due to their association with degradation phenomena such as leakage currents, polarization fatigue, and aging~\cite{yoo1992mechanism, tagantsev2001polarization, lupascu2006aging}. However, recent studies have begun to explore the potential benefits of controlled defect engineering in ferroelectrics. Introducing and managing specific defects can significantly influence material characteristics, including electrical conductivity, polarization stability, carrier dynamics, and magnetic ordering~\cite{walsh2017instilling, Ricca2022}.

Antisite defects are a type of point defect that occurs in materials with multiple cations on distinct sublattices, where one cation occupies the lattice site of the other~\cite{Ricca2022}. These defects tend to form under non-stoichiometric conditions, where an imbalance in cation concentrations leads to an excess of one species and a deficiency of the other. Their formation is influenced by factors such as ionic radii, induced lattice strain, and the relative charge of the involved cations. These kind of defects were shown to induce ferroelectricity in materials such as \ce{SrTiO3}~\cite{klyukin2017effect}, \ce{PbZrO3}~\cite{gao2017ferroelectricity} and \ce{YFeO3}~\cite{ning2021antisite}. While \ce{SrTiO3} and \ce{PbZrO3} are non-magnetic, antisite formation in \ce{YFeO3} results in a \ce{Y}-rich composition, reducing magnetic order and potentially weakening its multiferroic properties.

\begin{figure}
	\centering
	\includegraphics[width=0.8\columnwidth]{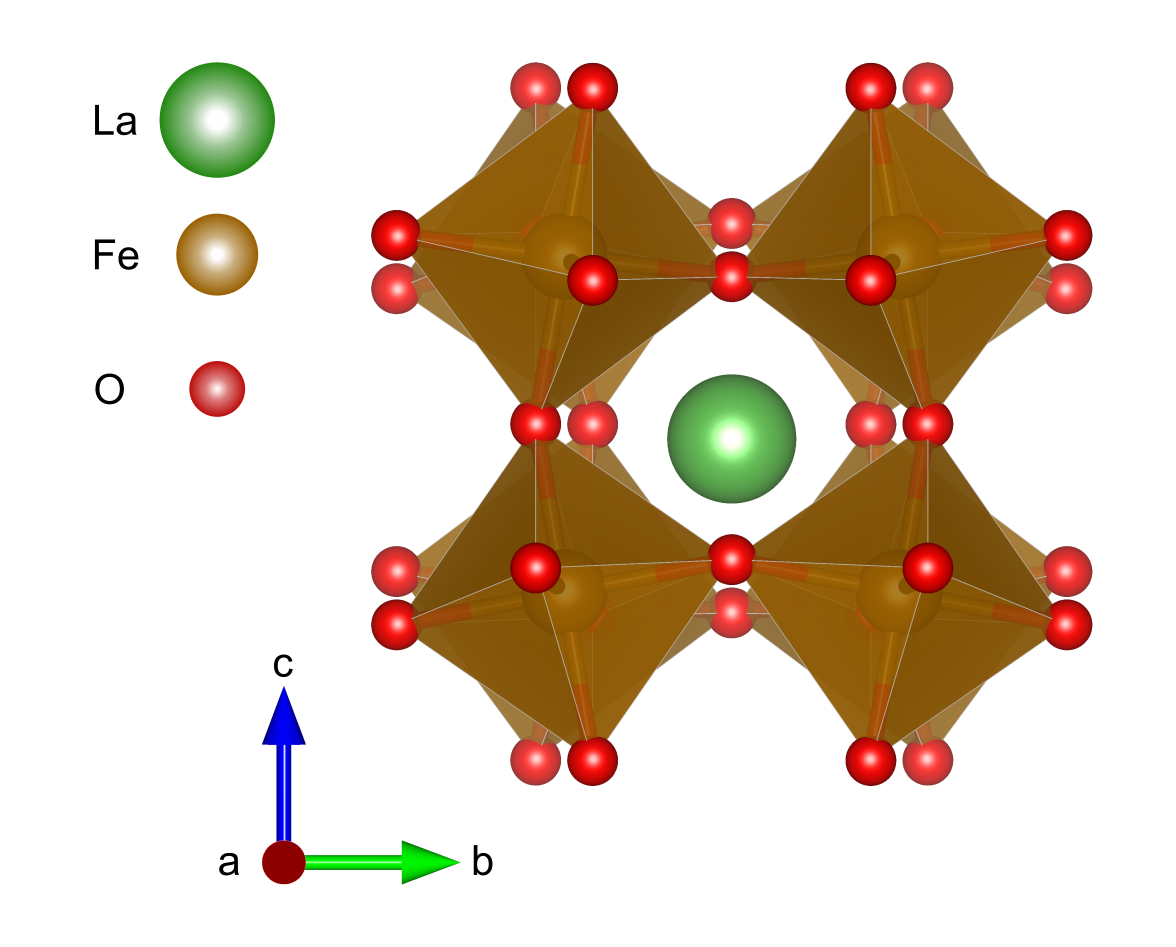}
	\caption{Structure of stoichiometric \ce{LaFeO3}. The antisite will be formed by substitution of a Fe on the central La atom.}
	\label{fig:struct}
\end{figure}

\ce{LaFeO3} (see Fig.~\ref{fig:struct}) is a non-polar orthoferrite with the ${Pbnm}$ space group, adopting a G-type antiferromagnetic (AFM) order with a N{\'e}el temperature of 740 K, highest among the rare-earth orthoferrites family~\cite{eibschutz1967mossbauer}. This structure is characterized by a ${a^{-}b^{+}a^{-}}$ octahedral tilting pattern in Glazer notation~\cite{glazer1972classification}, denoting equal out-of-phase tilts along the $a$ and $c$ axes, and in-phase tilts along the $b$ axis.

Here we explore a pathway to induce ferroelectricity into \ce{LaFeO3} via \ce{Fe_{La}} antisite defects. Our results demonstrate that these antisite defect disrupts the centrosymmetry of the lattice, creating a polar state that enables switchable ferroelectric behavior. Moreover, this defect changes the antiferromagnetic to a ferrimagnetic order, the polarization of which couples to the ferroelectric polarization. While on average cancelling in the pristine state, engineering these electric and magnetic defects dipoles under applied fields leads to net magnetoelectric coupling. This discovery opens a new route for achieving ferroelectricity and magnetoelectric coupling and highlights the potential of defect engineering to precisely tailor material properties for advanced electronic applications.

%%%%%%%%%%%%%%%%%%%%%%%%%%%%%%%%%%%%%%%%%%%%%%%%%%%%%%%%%%%%%%%%%%%%%%%%%%
\section{Methods}
%%%%%%%%%%%%%%%%%%%%%%%%%%%%%%%%%%%%%%%%%%%%%%%%%%%%%%%%%%%%%%%%%%%%%%%%%%

We performed density functional theory (DFT) calculations using the Vienna Ab initio Simulation Package (VASP)~\cite{Kresse:1993ty, Kresse:1994us, Kresse:1996vk, Kresse:1996vf}. The PBEsol exchange-correlation functional was employed with a Hubbard $U$ of 5.3 eV applied to \ce{Fe} 3d orbitals~\cite{perdew2008restoring, anisimov1991band, Dudarev1998}. We used projector augmented-wave (PAW) potentials with \ce{La} (5s, 5p, 5d, 6s), \ce{Fe} (3s, 3p, 4s, 3d), and \ce{O} (2s, 2p) valence electrons~\cite{blochl1994projector, kresse1999ultrasoft}. Wave functions were expanded in a plane-wave basis with a kinetic energy cutoff of 650 eV. Brillouin zone sampling was performed using a $4\times4\times4$ Monkhorst-Pack~\cite{monkhorst1976} $k$-point mesh for the 40 atom supercell.

Structures were relaxed until Hellmann-Feynman forces and the stress tensor converged below 1 meV/\AA\ and 0.08 meV/\AA$^3$ respectively. This structural optimization yielded lattice parameters $a = 7.833$, $b = 7.814$, and $c = 7.833$ \AA\ for the pristine 40-atom pseudo-cubic \ce{LaFeO3} cell. For defective supercells, we fixed the lattice vectors to these optimized stoichiometric values and relaxed only the internal atomic coordinates.

We calculated polarizations using Born effective charges obtained via density functional perturbation theory within the modern theory of polarization~\cite{gajdos2006linear, baroni1986dielectric} for the centrosymmetric structure. In this approach, the spontaneous polarization is obtained by multiplying the ionic displacements between the polar and a reference centrosymmetric structure with the Born effective charge tensors~\cite{spaldin2012beginner}. Finally, the polarization switching barrier $E_b$ was determined using the climbing image nudged elastic band (NEB) method~\cite{henkelman2000climbing}.

%%%%%%%%%%%%%%%%%%%%%%%%%%%%%%%%%%%%%%%%%%%%%%%%%%%%%%%%%%%%%%%%%%%%%%%%%%
\section{Results and Discussion}
%%%%%%%%%%%%%%%%%%%%%%%%%%%%%%%%%%%%%%%%%%%%%%%%%%%%%%%%%%%%%%%%%%%%%%%%%%

To evaluate the thermodynamic stability of \ce{Fe_{La}} antisite defects, we calculate their formation energy according to the formalism of Zhang and Northrup~\cite{zhang1991chemical,freysoldt2014first}:
\begin{align}
	E_\mathrm{form} &= E_{\mathrm{defect}} - E_{\mathrm{bulk}} - \sum_i n_i \mu_i \notag\\ &+ q(E_\mathrm{Fermi} + E_{\mathrm{VBM}}) + \Delta E_{\mathrm{corr}},
\end{align}
where $E_{\text{defect}}$ and $E_{\text{bulk}}$ are the DFT energies of the defective and stoichiometric cells respectively, $n_i$  denotes the number of added ($n_i>0$) or removed ($n_i<0$) atoms of element $i$ at chemical potential $\mu_i$. To account for different charge states $q$ of the defect, electrons are exchanged with a reservoir at $E_\mathrm{Fermi}$, which is taken relative to the valence band maximum ($E_\mathrm{VBM}$), and $\Delta E_{\mathrm{corr}}$ is a corrective term to align the electrostatic potentials of neutral and charged cells and account for image-charge interaction~\cite{lany2009accurate}.
\begin{figure}
	\centering
	\includegraphics[width=\columnwidth]{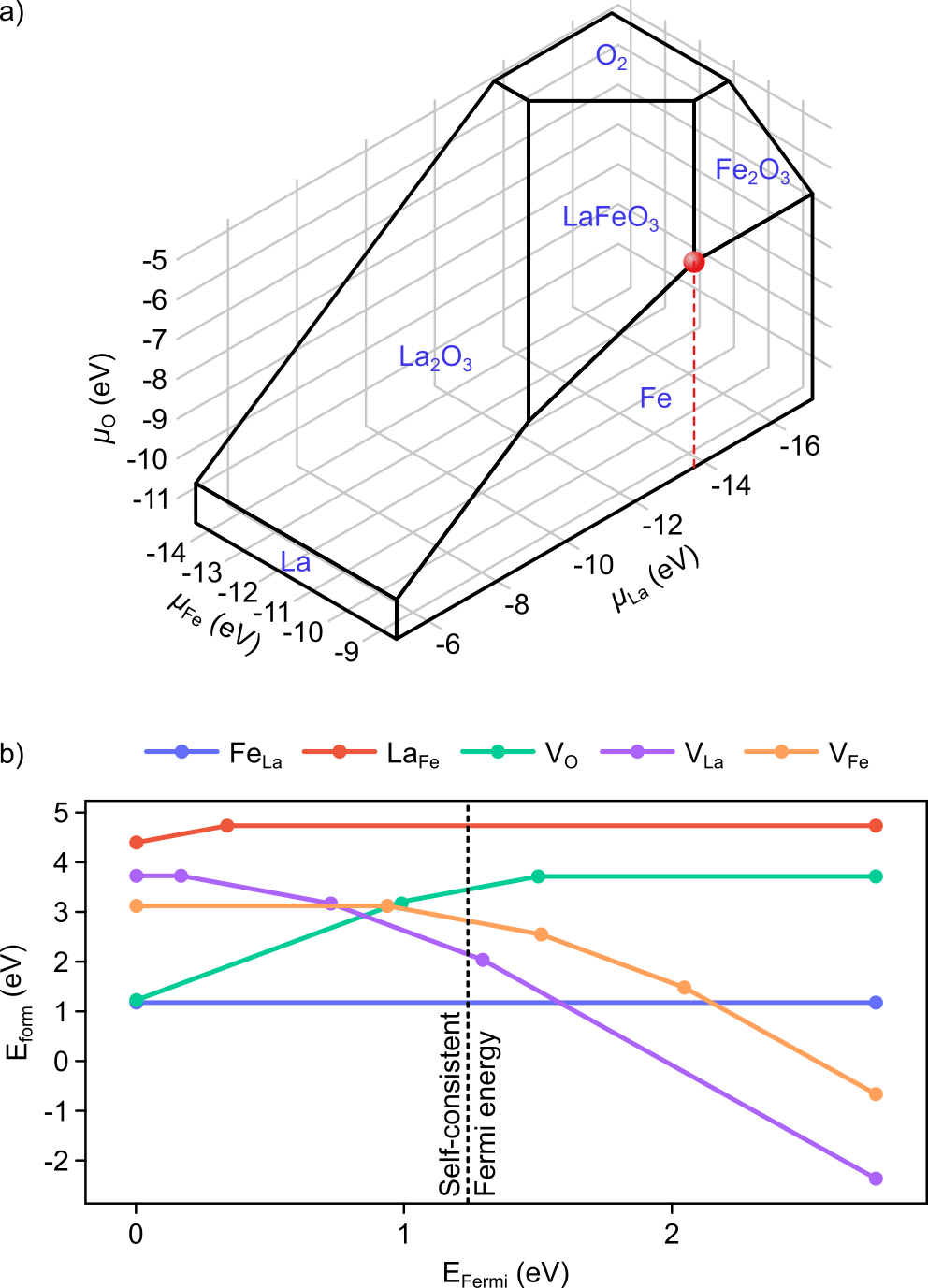}
	\caption{a) computed chemical potential diagram of the La-Fe-O system, showing the stability range of \ce{LaFeO3} and potential competing phases. The red dot shows the selected La-poor and Fe-rich conditions. b) computed formation energies of intrinsic antisite and vacancy defects as function of the Fermi energy. The dashed vertical line denotes the self-consistent Fermi energy \ce{LaFeO3} will assume with a defect population stable under the selected conditions.}
	\label{fig:phaseform}
\end{figure}

A \ce{Fe_{La}} antisite defect is expected to be most stable under La-poor and Fe-rich conditions. We use pymatgen~\cite{ong2013pymatgen} to determine the chemical potential ranges in which \ce{LaFeO3} is stable (see Fig.~\ref{fig:phaseform}a) and select the most La-poor and Fe-rich conditions, which correspond to $\mu_{\ce{La}} = -13.56$~eV and $\mu_{\ce{Fe}} = -8.37$~eV. We then use py-sc-fermi~\cite{pyscfermi} to determine the self-consistent Fermi energy, where the neutral \ce{Fe_{La}} has the lowest formation energy, approximately 1.0~eV lower than that of competing defects such as \ce{V_{La}}, \ce{V_{O}}, or \ce{La_{Fe}}. The magnitude of formation energies of the \ce{Fe_{La}} antisite and the oxygen vacancy are in good agreement with results obtained by quantum Monte Carlo~\cite{ichibha2023existence}. This \ce{Fe_{La}} formation energy of 1.1~eV, results in a defect concentration of $3.2\cdot10^{16}$~cm$^{-3}$ (0.0002~\%) at typical PLD temperatures of around 1000~K and $6.6\cdot10^{17}$~cm$^{-3}$ (0.004~\%) at typical solid-state synthesis temperatures of around 1300~K.

\begin{figure}
	\centering
	\includegraphics[width=\columnwidth]{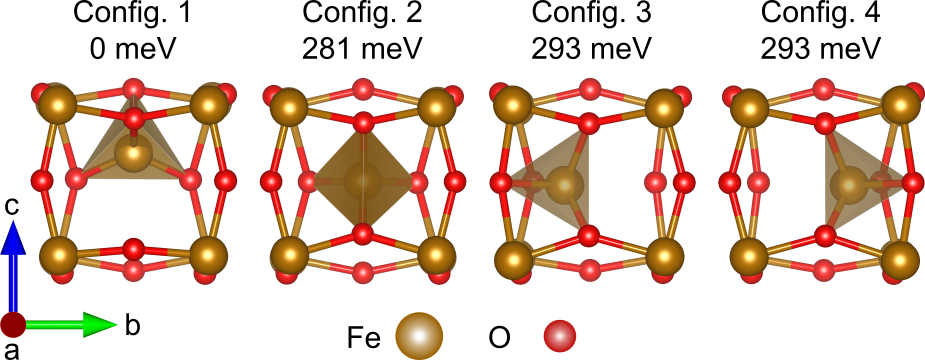}
	\caption{Most stable configurations of the \ce{Fe_{La}} antisite defect, along with their relative energy. In all cases the \ce{Fe} displaces from the center of the coordination polyhedron, assuming a tetrahedral coordination.}
	\label{fig:LaFeO3}
\end{figure}

As the radius of \ce{Fe} (Shannon~\cite{shannon1976revised} radius of 0.49-0.78~\AA\ depending on the coordination) is much smaller than that of \ce{La} (1.66~\AA\ in dodecahedral coordination), it seems natural for \ce{Fe_{La}} to displace away from the high-symmetry position at the center of the coordination dodecahedron. To determine possible stable positions of \ce{Fe_{La}}, we have performed structural relaxations from 26 starting configurations obtained by displacing the \ce{Fe_{La}} by 0.4~\AA\ along all face, edge and corner directions of the pseudo-cubic cell. 

This sampling directly identifies three energetically favorable configurations by displacing the \ce{Fe_{La}} atom along $[001]$, $[0\bar{1}0]$ and $[010]$ directions (see Fig.~\ref{fig:LaFeO3}). In the most stable configuration 1, the interstitial is displaced by 1.048 \AA\ from the high-symmetry position, the \ce{Fe_{La}} assuming a tetrahedral coordination with \ce{Fe-O} bonds lengths in the range 1.93--1.98 \AA. In addition to this most stable configuration, we identify configurations 3 and 4 that are 293 meV higher in energy. A nudged elastic band (NEB) calculation between configurations 3 and 4 identified an additional configuration 2 with an energy between configuration 1 and configurations 3 and 4. 

\begin{figure}
 	\centering
 	\includegraphics[width=\columnwidth]{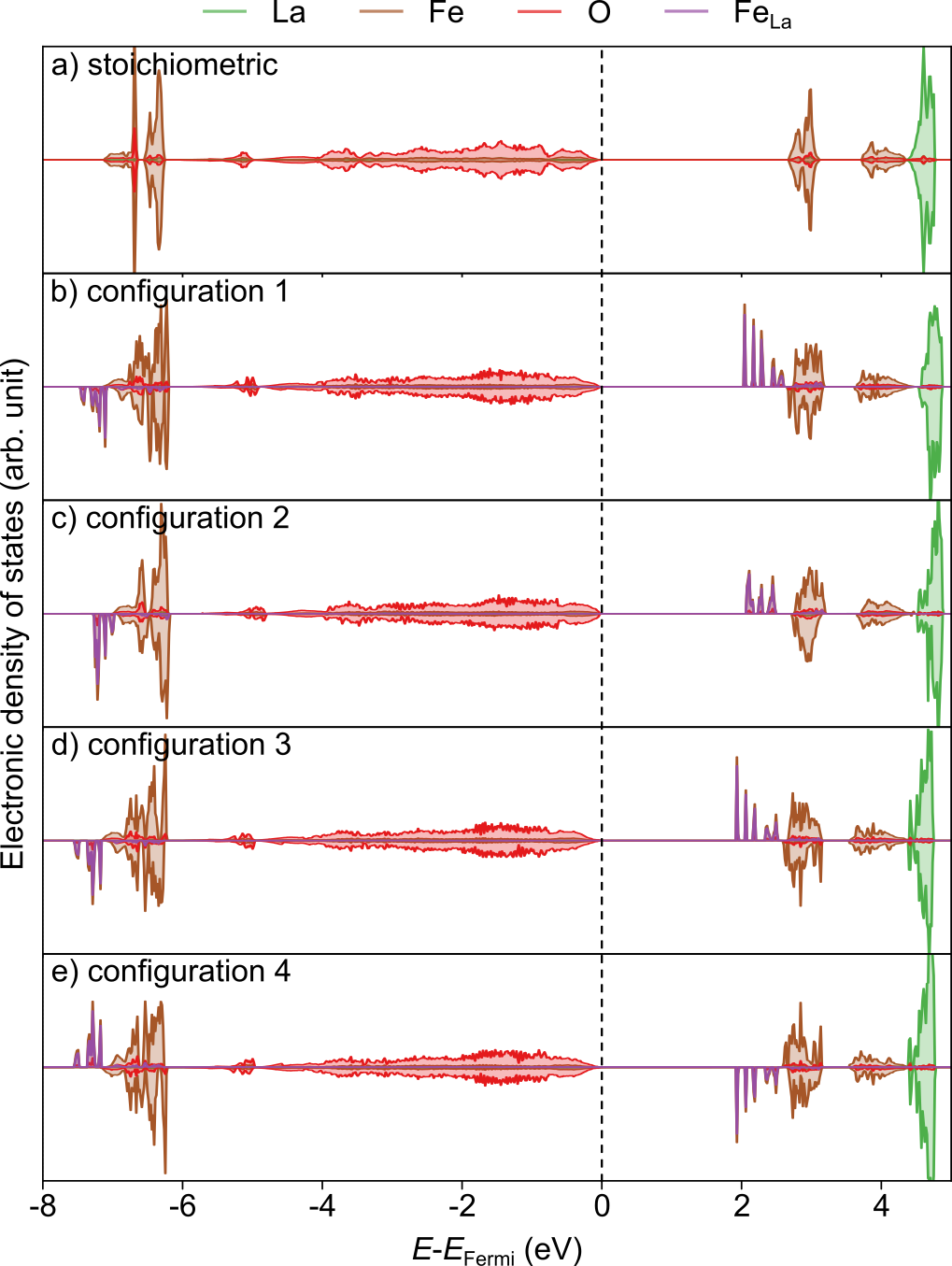}
 	\caption{Element-projected density of states for a) stoichiometric \ce{LaFeO3} and b-e) with \ce{Fe_{La}} antisite defects in configurations 1-4.}
 	\label{fig:DOS}
\end{figure}

In Fig.~\ref{fig:DOS}a), we show the element-projected density of states (DOS) for stoichiometric \ce{LaFeO3} and see an antiferromagnetic DOS with a semiconducting band gap of 2.4~eV typical for a charge-transfer insulator. Figs.~\ref{fig:DOS}b-e) show the DOS with \ce{Fe_{La}} antisites in each of the configurations identified above. We can see that the antisite introduces additional states at energies smaller than the lattice Fe-derived states around 6-7~eV below the Fermi energy. These states show very limited hybridization with oxygen, which is also the case for the unoccupied minority spin states that appear just below the conduction band edge. Since this alteration of the electronic structure is similar for all \ce{Fe_{La}} configurations, we can conclude that these defects impart a ferrimagnetic nature to \ce{LaFeO3}. Interestingly, we observe these antisite electrons in the down-spin channel for configurations 1--3, while they are in the up-spin channel for configuration 4. In fact, while for configurations 1 and 2, both antisite spin orientations are degenerate in energy, for configurations 3 and 4, the up/down configuration respectively would be 30~meV less favorable.

\begin{figure}
 	\centering
 	\includegraphics[width=\columnwidth]{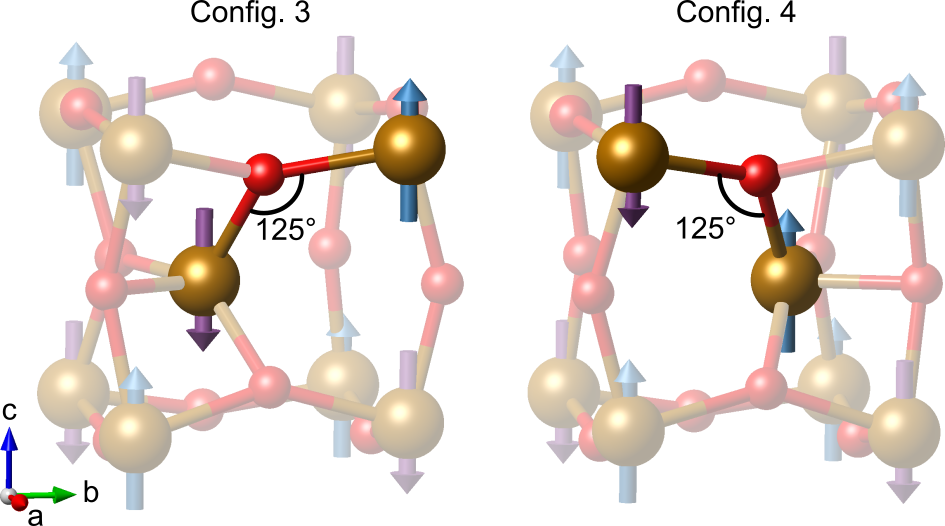}
	\caption{Dominant super-exchange pathways between a \ce{Fe_{La}} antisite and the G-type AFM lattice Fe atoms.}
	\label{fig:goodenough}
\end{figure}

This can be understood from the dominant super-exchange interactions in the two configurations. According to the Kanamori-Goodenough rules~\cite{goodenough1955theory, goodenough1958interpretation, kanamori1959superexchange} the larger the angle of a Fe--O--Fe bond, the stronger the antiferromagnetic super-exchange between the two Fe sites. As shown in Fig.~\ref{fig:goodenough} we find the largest angle in both configurations to be 125$^\circ$. Due to the G-type AFM order of \ce{LaFeO3}, this bond antiferromagnetically couples the antisite Fe with an up-spin Fe in configuration 3, while in configuration 4, it is with a down-spin Fe. The location of the antisite within the G-type AFM structure thus prefers one magnetic orientation of the antisite over the other. This picture is verified by the computed magnetic exchange constants as shown in the supporting information (SI) section~S2.

\begin{figure}
  \centering
  \includegraphics[width=\columnwidth]{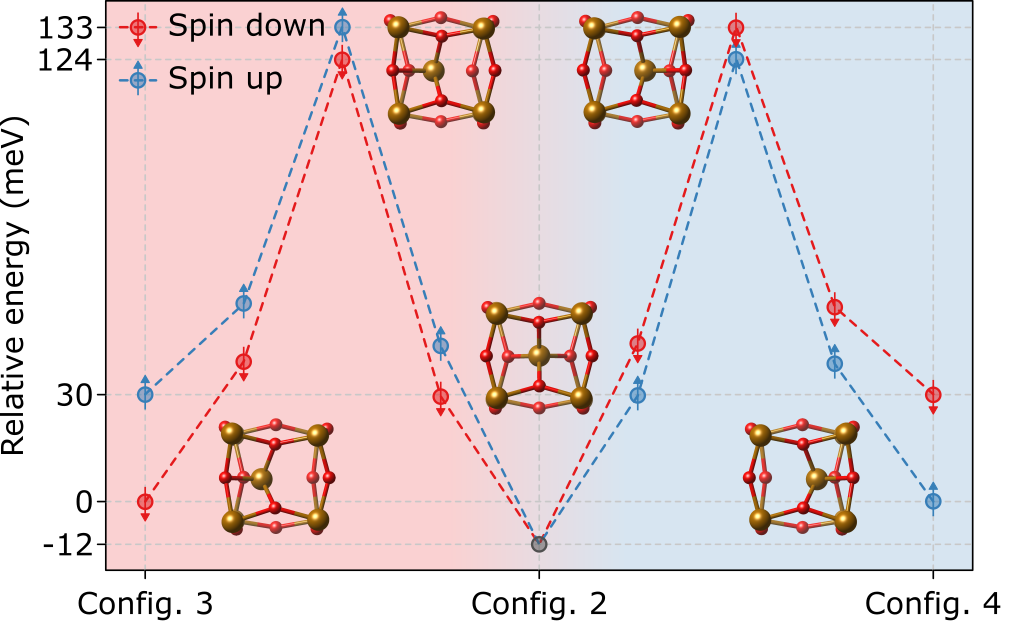}
  \caption{Nudged elastic band calculation between configurations 3 and 4, passing via the intermediate configuration 2. Atomic structures are shown along the $x$ axis, with the $y$ axis pointing right.}
  \label{fig:NEB}
\end{figure}

Given that configurations 3 and 4 are isoenergetic and mirror-symmetric with respect to the $(010)$ plane, they are possible states for ferroelectric switching, if the barrier for polarization reversal is sufficiently small and the polarization sufficiently large. To asses the magnitude of the barrier, we performed a nudged elastic band (NEB) calculation between configurations 3 and 4, the result of which is shown in Fig.~\ref{fig:NEB}. This path found the above-mentioned configuration 2 as an intermediate minimum, which was relaxed and at which the path was split. The resulting energy profile is thus the one of a triple-well ferroelectric with an energy barrier of 136 meV to switch from configuration 2 to the lower-energy magnetic state of either configuration 3 or 4, while the reverse process has a barrier of 124 meV. Our NEBs in the up and down spin magnetic state of the antisite show that the down spin antisite is more stable all the way from configuration 3 to 2, while the up spin antisite is more stable from configuration 2 to 4, the two being iso-energetic at configuration 2. This implies that along the switching pathway, the spin will likely switch at or around configuration 2 as shown by the shaded background in Fig.~\ref{fig:NEB}.

Since neither configurations 2, 3 or 4 are the ground state, we have also performed NEBs between the ground-state configuration 1 and both configurations 2 and 3, the results being shown in SI section~S3. We find that switching from configuration 1 to 2 has a barrier of 388 meV, while a transition from configuration 1 to 3 has a barrier of 389 meV. Importantly, the reverse switching is protected by barriers larger than 100 meV in both cases, which is significantly larger than $kT$ at room temperature.

\begin{figure}
  \centering
  \includegraphics[width=\columnwidth]{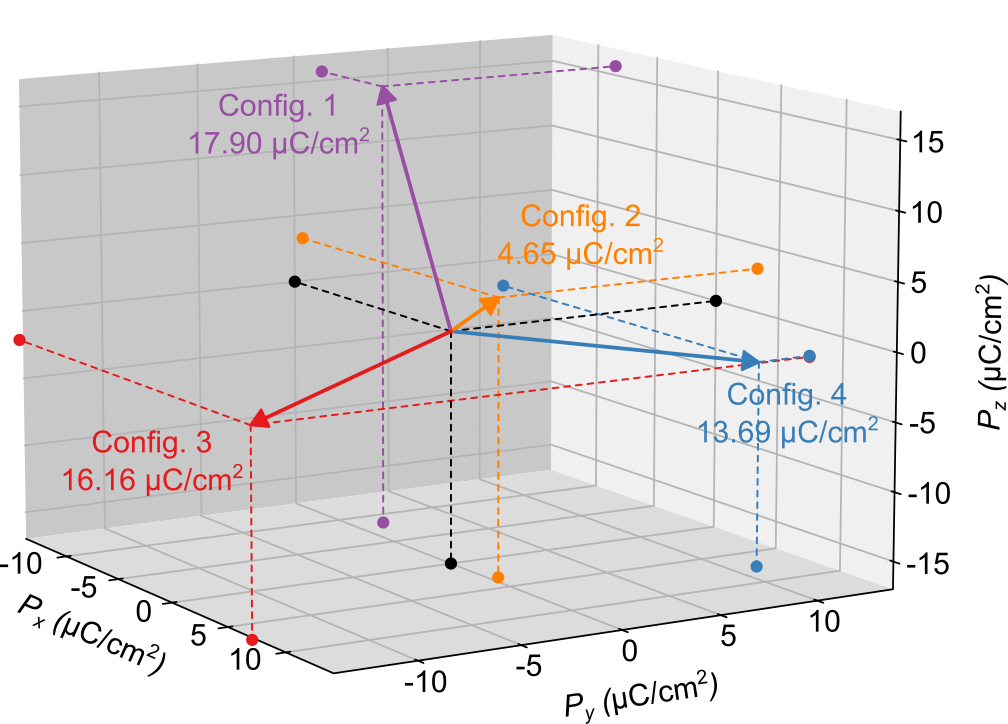}
  \caption{Polarization of configurations 1 to 4, calculated via the Born effective charge approach.}
  \label{fig:Polarization2D}
\end{figure}

Next, we compute the polarization of the different configurations via the Born effective charge (BEC) approach. As shown in Fig.~\ref{fig:Polarization2D}, we find sizable polarizations of 13.69 to 17.90 $\mu C/$cm$^2$ with main components along $x$ and $z$ for configuration 1 and $\mp y$ for configurations 3 and 4. Configuration 2 has a smaller polarization of 4.65 $\mu C/$cm$^2$ oriented primarily along $x$ and $z$. As shown in SI Table~S2 the displacement of the \ce{Fe_{La}} antisite accounts for about 60\% of the polarization, the rest coming from the relaxation of other atoms. The polarization of configurations 3 and 4 shows a slight asymmetry, which we ascribe to electronic effects captured by the BEC, as both configurations have equivalent polarizations in a point-charge model.

To assess the fields required for polarization switching, we computed the coercive field according to 
	\begin{align*} 	
		\epsilon_c = \frac{\left(\frac{4}{3}\right)^{3/2} E_b}{P V},
	\end{align*}
where $E_b$ is the switching energy barrier, $P$ is the polarization component along the switching direction (see SI section~S3) and $V$ is the volume of the supercell~\cite{rabe2007modern,beckman2009ideal}. As shown in SI Table~S3, we find coercive fields of around 400 MV/m for switching from either configuration 3 or 4 to 2, 1500 MV/m for the reverse switching, while switching from configuration 1 to 2 can be achieved with 1100 MV/m. We note that the volume appearing in the denominator will be larger than the simulation-cell volume we used for these estimations, since  defect concentrations should be lower than the ones computed here, leading to a marked reduction of the required fields. Therefore in the worst case, a film thickness/electrode spacing of around 5 nm should prevent Zener breakdown, given that \ce{LaFeO3} has a band gap of around 2.5 eV.

Based on these results, we propose that as-synthesized La-deficient and Fe-rich \ce{LaFeO3} will have \ce{Fe_{La}} antisites primarily in configuration 1. By applying an electric field along $x$, these antisites can be switched into configurations 2 or 3/4, at which point an electric field along $y$ can be used for ferroelectric switching. Falling back to configuration 1 is prevented by the reverse switching barrier that is larger than $kT$ at room temperature and the orthogonality of the electric fields. Nevertheless, a geometry with two sets of electrodes will be required for ferroelectric switching of antisites in \ce{LaFeO3}.

\begin{figure}
  \centering
  \includegraphics[width=\columnwidth]{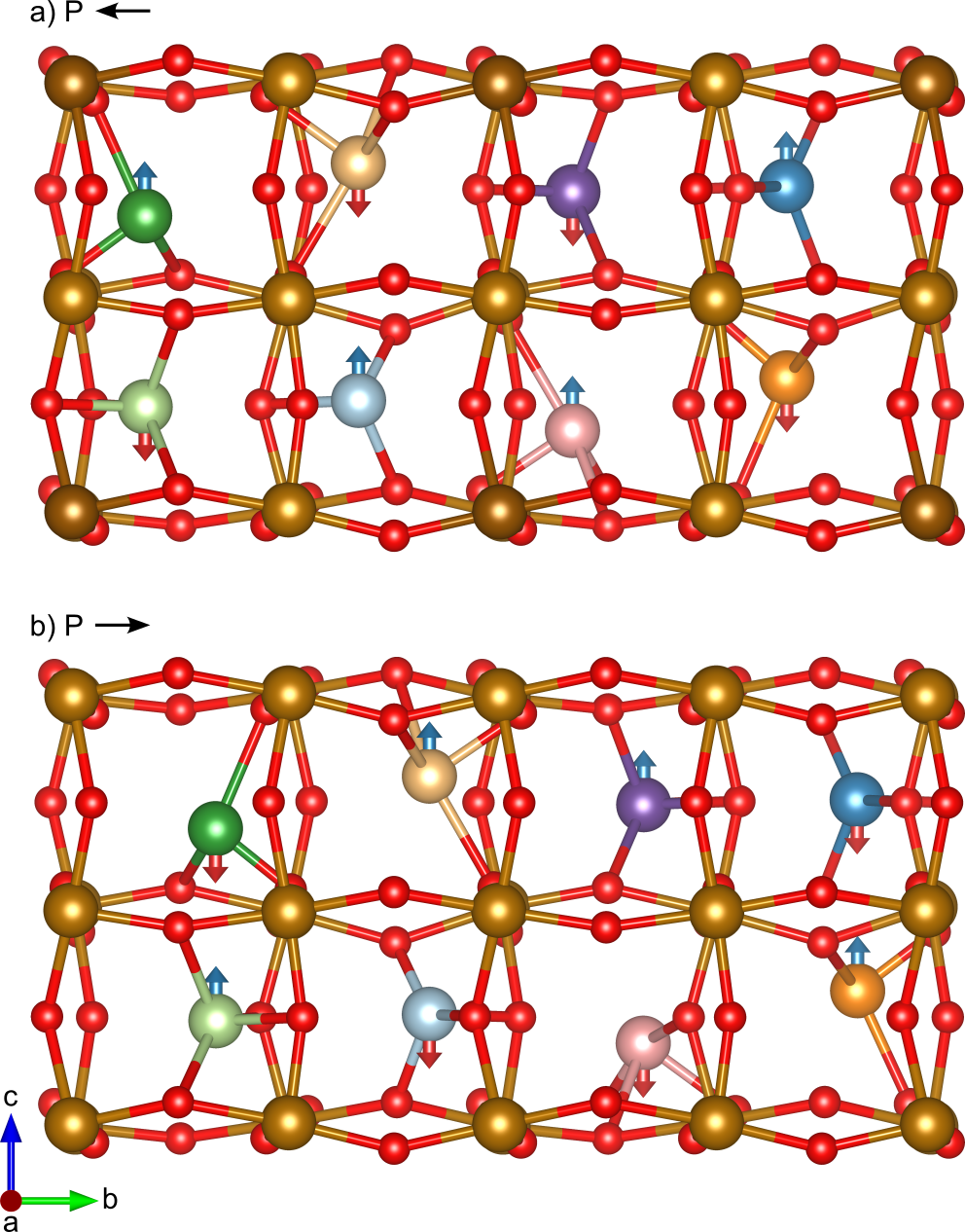}
  \caption{Schematic view of the preferred magnetic polarization of \ce{Fe_{La}} antisites in all A-site pockets \ce{LaFeO3} for a) a ferroelectric polarization pointing along $-y$ and b) a ferroelectric polarization pointing along $+y$.}
  \label{fig:All_Pockets}
\end{figure}

Up to now, we have shown that \ce{LaFeO3} with \ce{Fe_{La}} antisite defects is ferroelectric and that ferroelectric switching of a single antisite goes along with switching the magnetic orientation of this antisite, implying that a magnetoelectric coupling exists in this material. Since the antisites may form in any of the A-site pockets that have a different crystallographic environment, we next assess the behavior of an antisite in each of these pockets individually. Fig.~\ref{fig:All_Pockets} shows the results of these individual calculations combined into one figure. We can see that half of the pockets (dark green, brown, pink and orange), show a configuration resembling the one in Fig.~\ref{fig:LaFeO3}, while for others (light green, light blue, purple, dark blue) it looks visually different, but would be the same if seen along $c$. Importantly, all of the antisites show the above-mentioned spin reorientation when ferroelectricity is switched along the $y$ direction. We can, however, also see that half of the antisites show an up spin when the ferroelectric polarization points along $-y$, and the other half shown a down spin. Since we cannot control which pocket an antisite occupies, this will lead to an on-average spin-compensated (antiferrimagnetic) configuration.

We, however, see two avenues to make a material with these properties interesting for applications. For the first, we assume \ce{LaFeO3} with antisites held at the same time in a magnetic field that will favor the parallel alignment of antisite spins and in an electric field along $y$ that will favor one ferroelectric polarization direction. Due to the low magneto-crystalline anisotropy of only 33 $\mu$eV for \ce{LaFeO3}~\cite{chen2025pure}, 180$^\circ$ switching of the G-type AFM order around antisites of the non-favorable spin orientation may take place. After removing the magnetic field, the result is a structure with AFM domains around each of the antisites and the formation of domain walls, the energy of which should be fairly low, again due to the low magneto-crystalline anisotropy. In this multi-domain configuration, all antisite spins point in the same direction, reverse upon polarization switching, leading to a net ferroelectric-ferrimagnetic magnetoelectric coupling.

The second application also uses the ferrimagnetic impurity in the AFM lattice. As discussed above, operating the material under an applied magnetic field can lead to 180$^\circ$ switching of the Néel vector of the lattice AFM order to optimize the exchange interactions with the antisite. Switching the ferroelectric polarization under applied magnetic field, which clamps the antisite spin, can therefore lead to electrically induced 180$^\circ$ switching of the AFM lattice. We note that the magnetic field could be static rather than generated via an electric current.

%%%%%%%%%%%%%%%%%%%%%%%%%%%%%%%%%%%%%%%%%%%%%%%%%%%%%%%%%%%%%%%%%%%%%%%%%%%%%%%%%
\section{Conclusions}
%%%%%%%%%%%%%%%%%%%%%%%%%%%%%%%%%%%%%%%%%%%%%%%%%%%%%%%%%%%%%%%%%%%%%%%%%%%%%%%%%

We study \ce{Fe_{La}} antisite defects in \ce{LaFeO3} by density functional theory calculations and find these defects to be favorable under La-poor and Fe-rich conditions. Due to the ionic size difference, we find different configurations for the antisite within the A-site pocket that induce switchable ferroelectric polarization. In order to reach a state with opposite polarizations along the $y$ axis, the material needs to be activated by an electric field along $x$, after which fields applied along $y$ can switch the polarization.

The antisite also induces an additional spin, which renders the antiferromagnetic \ce{LaFeO3} ferrimagnetic. Excitingly, we predict a switching of this antisite spin with ferroelectric switching, implying magnetoelectric coupling in the material. On average, we predict the antisite spins to cancel in different A-site pockets. However, preparing the material under an applied magnetic field could lead to antiferromagnetic domain walls and a net ferroelectric-ferrimagnetic coupling. On the other hand ferroelectric switching under a static magnetic field can lead to 180$^\circ$ switching of the antiferromagnetic order. Both these application pathways hold great promise for data-storage applications.

%%%%%%%%%%%%%%%%%%%%%%%%%%%%%%%%%%%%%%%%%%%%%%%%%%%%%%%%%%%%%%%%%%%%%%%%%%%%%%%%%
\bibliography{references}

\begin{thebibliography}{42}%
\makeatletter
\providecommand \@ifxundefined [1]{%
 \@ifx{#1\undefined}
}%
\providecommand \@ifnum [1]{%
 \ifnum #1\expandafter \@firstoftwo
 \else \expandafter \@secondoftwo
 \fi
}%
\providecommand \@ifx [1]{%
 \ifx #1\expandafter \@firstoftwo
 \else \expandafter \@secondoftwo
 \fi
}%
\providecommand \natexlab [1]{#1}%
\providecommand \enquote  [1]{``#1''}%
\providecommand \bibnamefont  [1]{#1}%
\providecommand \bibfnamefont [1]{#1}%
\providecommand \citenamefont [1]{#1}%
\providecommand \href@noop [0]{\@secondoftwo}%
\providecommand \href [0]{\begingroup \@sanitize@url \@href}%
\providecommand \@href[1]{\@@startlink{#1}\@@href}%
\providecommand \@@href[1]{\endgroup#1\@@endlink}%
\providecommand \@sanitize@url [0]{\catcode `\\12\catcode `\$12\catcode `\&12\catcode `\#12\catcode `\^12\catcode `\_12\catcode `\%12\relax}%
\providecommand \@@startlink[1]{}%
\providecommand \@@endlink[0]{}%
\providecommand \url  [0]{\begingroup\@sanitize@url \@url }%
\providecommand \@url [1]{\endgroup\@href {#1}{\urlprefix }}%
\providecommand \urlprefix  [0]{URL }%
\providecommand \Eprint [0]{\href }%
\providecommand \doibase [0]{http://dx.doi.org/}%
\providecommand \selectlanguage [0]{\@gobble}%
\providecommand \bibinfo  [0]{\@secondoftwo}%
\providecommand \bibfield  [0]{\@secondoftwo}%
\providecommand \translation [1]{[#1]}%
\providecommand \BibitemOpen [0]{}%
\providecommand \bibitemStop [0]{}%
\providecommand \bibitemNoStop [0]{.\EOS\space}%
\providecommand \EOS [0]{\spacefactor3000\relax}%
\providecommand \BibitemShut  [1]{\csname bibitem#1\endcsname}%
\let\auto@bib@innerbib\@empty
%</preamble>
\bibitem [{\citenamefont {Son}\ \emph {et~al.}(2013)\citenamefont {Son}, \citenamefont {Lee}, \citenamefont {Song}, \citenamefont {Shin},\ and\ \citenamefont {Jang}}]{son2013four}%
  \BibitemOpen
  \bibfield  {author} {\bibinfo {author} {\bibfnamefont {J.~Y.}\ \bibnamefont {Son}}, \bibinfo {author} {\bibfnamefont {J.-H.}\ \bibnamefont {Lee}}, \bibinfo {author} {\bibfnamefont {S.}~\bibnamefont {Song}}, \bibinfo {author} {\bibfnamefont {Y.-H.}\ \bibnamefont {Shin}}, \ and\ \bibinfo {author} {\bibfnamefont {H.~M.}\ \bibnamefont {Jang}},\ }\href {\doibase 10.1021/nn4017422} {\bibfield  {journal} {\bibinfo  {journal} {ACS Nano}\ }\textbf {\bibinfo {volume} {7}},\ \bibinfo {pages} {5522} (\bibinfo {year} {2013})}\BibitemShut {NoStop}%
\bibitem [{\citenamefont {Ruan}\ \emph {et~al.}(2016)\citenamefont {Ruan}, \citenamefont {Li}, \citenamefont {Yuan}, \citenamefont {Wang}, \citenamefont {Li},\ and\ \citenamefont {Wu}}]{ruan2016four}%
  \BibitemOpen
  \bibfield  {author} {\bibinfo {author} {\bibfnamefont {J.}~\bibnamefont {Ruan}}, \bibinfo {author} {\bibfnamefont {C.}~\bibnamefont {Li}}, \bibinfo {author} {\bibfnamefont {Z.}~\bibnamefont {Yuan}}, \bibinfo {author} {\bibfnamefont {P.}~\bibnamefont {Wang}}, \bibinfo {author} {\bibfnamefont {A.}~\bibnamefont {Li}}, \ and\ \bibinfo {author} {\bibfnamefont {D.}~\bibnamefont {Wu}},\ }\href {\doibase 10.1063/1.4972786} {\bibfield  {journal} {\bibinfo  {journal} {Appl. Phys. Lett.}\ }\textbf {\bibinfo {volume} {109}},\ \bibinfo {pages} {252903} (\bibinfo {year} {2016})}\BibitemShut {NoStop}%
\bibitem [{\citenamefont {Ramesh}\ and\ \citenamefont {Manipatruni}(2021)}]{ramesh2021electric}%
  \BibitemOpen
  \bibfield  {author} {\bibinfo {author} {\bibfnamefont {R.}~\bibnamefont {Ramesh}}\ and\ \bibinfo {author} {\bibfnamefont {S.}~\bibnamefont {Manipatruni}},\ }\href {\doibase 10.1098/rspa.2020.0942} {\bibfield  {journal} {\bibinfo  {journal} {Proc. R. Soc. A}\ }\textbf {\bibinfo {volume} {477}},\ \bibinfo {pages} {20200942} (\bibinfo {year} {2021})}\BibitemShut {NoStop}%
\bibitem [{\citenamefont {Li}\ \emph {et~al.}(2023)\citenamefont {Li}, \citenamefont {Singh}, \citenamefont {Bao}, \citenamefont {Luo}, \citenamefont {Li}, \citenamefont {Chatterjee}, \citenamefont {Goiriena-Goikoetxea}, \citenamefont {Xiao}, \citenamefont {Tamura}, \citenamefont {Candler}, \citenamefont {You}, \citenamefont {Bokor},\ and\ \citenamefont {Hong}}]{li2023energy}%
  \BibitemOpen
  \bibfield  {author} {\bibinfo {author} {\bibfnamefont {X.}~\bibnamefont {Li}}, \bibinfo {author} {\bibfnamefont {H.}~\bibnamefont {Singh}}, \bibinfo {author} {\bibfnamefont {Y.}~\bibnamefont {Bao}}, \bibinfo {author} {\bibfnamefont {Q.}~\bibnamefont {Luo}}, \bibinfo {author} {\bibfnamefont {S.}~\bibnamefont {Li}}, \bibinfo {author} {\bibfnamefont {J.}~\bibnamefont {Chatterjee}}, \bibinfo {author} {\bibfnamefont {M.}~\bibnamefont {Goiriena-Goikoetxea}}, \bibinfo {author} {\bibfnamefont {Z.}~\bibnamefont {Xiao}}, \bibinfo {author} {\bibfnamefont {N.}~\bibnamefont {Tamura}}, \bibinfo {author} {\bibfnamefont {R.~N.}\ \bibnamefont {Candler}}, \bibinfo {author} {\bibfnamefont {L.}~\bibnamefont {You}}, \bibinfo {author} {\bibfnamefont {J.}~\bibnamefont {Bokor}}, \ and\ \bibinfo {author} {\bibfnamefont {J.}~\bibnamefont {Hong}},\ }\href {\doibase 10.1021/acs.nanolett.3c00707} {\bibfield  {journal} {\bibinfo  {journal} {Nano Lett.}\ }\textbf {\bibinfo {volume} {23}},\ \bibinfo {pages} {6845} (\bibinfo {year} {2023})}\BibitemShut {NoStop}%
\bibitem [{\citenamefont {Pe{\~n}a}\ and\ \citenamefont {Fierro}(2001)}]{pena2001chemical}%
  \BibitemOpen
  \bibfield  {author} {\bibinfo {author} {\bibfnamefont {M.}~\bibnamefont {Pe{\~n}a}}\ and\ \bibinfo {author} {\bibfnamefont {J.}~\bibnamefont {Fierro}},\ }\href {\doibase 10.1021/cr980129f} {\bibfield  {journal} {\bibinfo  {journal} {Chem. Rev.}\ }\textbf {\bibinfo {volume} {101}},\ \bibinfo {pages} {1981} (\bibinfo {year} {2001})}\BibitemShut {NoStop}%
\bibitem [{\citenamefont {Ricca}\ and\ \citenamefont {Aschauer}(2022)}]{Ricca2022}%
  \BibitemOpen
  \bibfield  {author} {\bibinfo {author} {\bibfnamefont {C.}~\bibnamefont {Ricca}}\ and\ \bibinfo {author} {\bibfnamefont {U.}~\bibnamefont {Aschauer}},\ }\href {\doibase 10.1007/s00339-022-06210-8} {\bibfield  {journal} {\bibinfo  {journal} {Appl. Phys. A}\ }\textbf {\bibinfo {volume} {128}},\ \bibinfo {pages} {1083} (\bibinfo {year} {2022})}\BibitemShut {NoStop}%
\bibitem [{\citenamefont {Yoo}\ and\ \citenamefont {Desu}(1992)}]{yoo1992mechanism}%
  \BibitemOpen
  \bibfield  {author} {\bibinfo {author} {\bibfnamefont {I.-K.}\ \bibnamefont {Yoo}}\ and\ \bibinfo {author} {\bibfnamefont {S.}~\bibnamefont {Desu}},\ }\href {\doibase 10.1002/pssa.2211330242} {\bibfield  {journal} {\bibinfo  {journal} {Phys. Status Solidi A}\ }\textbf {\bibinfo {volume} {133}},\ \bibinfo {pages} {565} (\bibinfo {year} {1992})}\BibitemShut {NoStop}%
\bibitem [{\citenamefont {Tagantsev}\ \emph {et~al.}(2001)\citenamefont {Tagantsev}, \citenamefont {Stolichnov}, \citenamefont {Colla},\ and\ \citenamefont {Setter}}]{tagantsev2001polarization}%
  \BibitemOpen
  \bibfield  {author} {\bibinfo {author} {\bibfnamefont {A.~K.}\ \bibnamefont {Tagantsev}}, \bibinfo {author} {\bibfnamefont {I.}~\bibnamefont {Stolichnov}}, \bibinfo {author} {\bibfnamefont {E.}~\bibnamefont {Colla}}, \ and\ \bibinfo {author} {\bibfnamefont {N.}~\bibnamefont {Setter}},\ }\href {\doibase 10.1063/1.1381542} {\bibfield  {journal} {\bibinfo  {journal} {J. Appl. Phys.}\ }\textbf {\bibinfo {volume} {90}},\ \bibinfo {pages} {1387} (\bibinfo {year} {2001})}\BibitemShut {NoStop}%
\bibitem [{\citenamefont {Lupascu}\ \emph {et~al.}(2006)\citenamefont {Lupascu}, \citenamefont {Genenko},\ and\ \citenamefont {Balke}}]{lupascu2006aging}%
  \BibitemOpen
  \bibfield  {author} {\bibinfo {author} {\bibfnamefont {D.~C.}\ \bibnamefont {Lupascu}}, \bibinfo {author} {\bibfnamefont {Y.~A.}\ \bibnamefont {Genenko}}, \ and\ \bibinfo {author} {\bibfnamefont {N.}~\bibnamefont {Balke}},\ }\href {\doibase 10.1111/j.1551-2916.2005.00663.x} {\bibfield  {journal} {\bibinfo  {journal} {J. Am. Ceram. Soc.}\ }\textbf {\bibinfo {volume} {89}},\ \bibinfo {pages} {224} (\bibinfo {year} {2006})}\BibitemShut {NoStop}%
\bibitem [{\citenamefont {Walsh}\ and\ \citenamefont {Zunger}(2017)}]{walsh2017instilling}%
  \BibitemOpen
  \bibfield  {author} {\bibinfo {author} {\bibfnamefont {A.}~\bibnamefont {Walsh}}\ and\ \bibinfo {author} {\bibfnamefont {A.}~\bibnamefont {Zunger}},\ }\href {\doibase 10.1038/nmat4973} {\bibfield  {journal} {\bibinfo  {journal} {Nat. Mat.}\ }\textbf {\bibinfo {volume} {16}},\ \bibinfo {pages} {964} (\bibinfo {year} {2017})}\BibitemShut {NoStop}%
\bibitem [{\citenamefont {Klyukin}\ and\ \citenamefont {Alexandrov}(2017)}]{klyukin2017effect}%
  \BibitemOpen
  \bibfield  {author} {\bibinfo {author} {\bibfnamefont {K.}~\bibnamefont {Klyukin}}\ and\ \bibinfo {author} {\bibfnamefont {V.}~\bibnamefont {Alexandrov}},\ }\href {\doibase 10.1103/PhysRevB.95.035301} {\bibfield  {journal} {\bibinfo  {journal} {Phys. Rev. B}\ }\textbf {\bibinfo {volume} {95}},\ \bibinfo {pages} {035301} (\bibinfo {year} {2017})}\BibitemShut {NoStop}%
\bibitem [{\citenamefont {Gao}\ \emph {et~al.}(2017)\citenamefont {Gao}, \citenamefont {Reyes-Lillo}, \citenamefont {Xu}, \citenamefont {Dasgupta}, \citenamefont {Dong}, \citenamefont {Dedon}, \citenamefont {Kim}, \citenamefont {Saremi}, \citenamefont {Chen}, \citenamefont {Serrao}, \citenamefont {Zhou}, \citenamefont {Neaton},\ and\ \citenamefont {Martin}}]{gao2017ferroelectricity}%
  \BibitemOpen
  \bibfield  {author} {\bibinfo {author} {\bibfnamefont {R.}~\bibnamefont {Gao}}, \bibinfo {author} {\bibfnamefont {S.~E.}\ \bibnamefont {Reyes-Lillo}}, \bibinfo {author} {\bibfnamefont {R.}~\bibnamefont {Xu}}, \bibinfo {author} {\bibfnamefont {A.}~\bibnamefont {Dasgupta}}, \bibinfo {author} {\bibfnamefont {Y.}~\bibnamefont {Dong}}, \bibinfo {author} {\bibfnamefont {L.~R.}\ \bibnamefont {Dedon}}, \bibinfo {author} {\bibfnamefont {J.}~\bibnamefont {Kim}}, \bibinfo {author} {\bibfnamefont {S.}~\bibnamefont {Saremi}}, \bibinfo {author} {\bibfnamefont {Z.}~\bibnamefont {Chen}}, \bibinfo {author} {\bibfnamefont {C.~R.}\ \bibnamefont {Serrao}}, \bibinfo {author} {\bibfnamefont {H.}~\bibnamefont {Zhou}}, \bibinfo {author} {\bibfnamefont {J.~B.}\ \bibnamefont {Neaton}}, \ and\ \bibinfo {author} {\bibfnamefont {L.~W.}\ \bibnamefont {Martin}},\ }\href {\doibase 10.1021/acs.chemmater.7b02506} {\bibfield  {journal} {\bibinfo  {journal} {Chem. Mater.}\ }\textbf {\bibinfo {volume} {29}},\ \bibinfo {pages} {6544} (\bibinfo {year} {2017})}\BibitemShut {NoStop}%
\bibitem [{\citenamefont {Ning}\ \emph {et~al.}(2021)\citenamefont {Ning}, \citenamefont {Kumar}, \citenamefont {Klyukin}, \citenamefont {Cho}, \citenamefont {Kim}, \citenamefont {Su}, \citenamefont {Kim}, \citenamefont {LeBeau}, \citenamefont {Yildiz},\ and\ \citenamefont {Ross}}]{ning2021antisite}%
  \BibitemOpen
  \bibfield  {author} {\bibinfo {author} {\bibfnamefont {S.}~\bibnamefont {Ning}}, \bibinfo {author} {\bibfnamefont {A.}~\bibnamefont {Kumar}}, \bibinfo {author} {\bibfnamefont {K.}~\bibnamefont {Klyukin}}, \bibinfo {author} {\bibfnamefont {E.}~\bibnamefont {Cho}}, \bibinfo {author} {\bibfnamefont {J.~H.}\ \bibnamefont {Kim}}, \bibinfo {author} {\bibfnamefont {T.}~\bibnamefont {Su}}, \bibinfo {author} {\bibfnamefont {H.-S.}\ \bibnamefont {Kim}}, \bibinfo {author} {\bibfnamefont {J.~M.}\ \bibnamefont {LeBeau}}, \bibinfo {author} {\bibfnamefont {B.}~\bibnamefont {Yildiz}}, \ and\ \bibinfo {author} {\bibfnamefont {C.~A.}\ \bibnamefont {Ross}},\ }\href {\doibase 10.1038/s41467-021-24592-w} {\bibfield  {journal} {\bibinfo  {journal} {Nat. Commun.}\ }\textbf {\bibinfo {volume} {12}},\ \bibinfo {pages} {4298} (\bibinfo {year} {2021})}\BibitemShut {NoStop}%
\bibitem [{\citenamefont {Eibsch{\"u}tz}\ \emph {et~al.}(1967)\citenamefont {Eibsch{\"u}tz}, \citenamefont {Shtrikman},\ and\ \citenamefont {Treves}}]{eibschutz1967mossbauer}%
  \BibitemOpen
  \bibfield  {author} {\bibinfo {author} {\bibfnamefont {M.}~\bibnamefont {Eibsch{\"u}tz}}, \bibinfo {author} {\bibfnamefont {S.}~\bibnamefont {Shtrikman}}, \ and\ \bibinfo {author} {\bibfnamefont {D.}~\bibnamefont {Treves}},\ }\href {\doibase 10.1103/PhysRev.156.562} {\bibfield  {journal} {\bibinfo  {journal} {Phys. Rev.}\ }\textbf {\bibinfo {volume} {156}},\ \bibinfo {pages} {562} (\bibinfo {year} {1967})}\BibitemShut {NoStop}%
\bibitem [{\citenamefont {Glazer}(1972)}]{glazer1972classification}%
  \BibitemOpen
  \bibfield  {author} {\bibinfo {author} {\bibfnamefont {A.~M.}\ \bibnamefont {Glazer}},\ }\href {\doibase 10.1107/S0567740872007976} {\bibfield  {journal} {\bibinfo  {journal} {Acta Cryst. B}\ }\textbf {\bibinfo {volume} {28}},\ \bibinfo {pages} {3384} (\bibinfo {year} {1972})}\BibitemShut {NoStop}%
\bibitem [{\citenamefont {Kresse}\ and\ \citenamefont {Hafner}(1993)}]{Kresse:1993ty}%
  \BibitemOpen
  \bibfield  {author} {\bibinfo {author} {\bibfnamefont {G.}~\bibnamefont {Kresse}}\ and\ \bibinfo {author} {\bibfnamefont {J.}~\bibnamefont {Hafner}},\ }\href {\doibase 10.1103/physrevb.47.558} {\bibfield  {journal} {\bibinfo  {journal} {Phys. Rev. B}\ }\textbf {\bibinfo {volume} {47}},\ \bibinfo {pages} {558} (\bibinfo {year} {1993})}\BibitemShut {NoStop}%
\bibitem [{\citenamefont {Kresse}\ and\ \citenamefont {Hafner}(1994)}]{Kresse:1994us}%
  \BibitemOpen
  \bibfield  {author} {\bibinfo {author} {\bibfnamefont {G.}~\bibnamefont {Kresse}}\ and\ \bibinfo {author} {\bibfnamefont {J.}~\bibnamefont {Hafner}},\ }\href {\doibase 10.1103/physrevb.49.14251} {\bibfield  {journal} {\bibinfo  {journal} {Phys. Rev. B}\ }\textbf {\bibinfo {volume} {49}},\ \bibinfo {pages} {14251} (\bibinfo {year} {1994})}\BibitemShut {NoStop}%
\bibitem [{\citenamefont {Kresse}\ and\ \citenamefont {Furthm{\"u}ller}(1996{\natexlab{a}})}]{Kresse:1996vk}%
  \BibitemOpen
  \bibfield  {author} {\bibinfo {author} {\bibfnamefont {G.}~\bibnamefont {Kresse}}\ and\ \bibinfo {author} {\bibfnamefont {J.}~\bibnamefont {Furthm{\"u}ller}},\ }\href {\doibase 10.1016/0927-0256(96)00008-0} {\bibfield  {journal} {\bibinfo  {journal} {Comput. Mater. Sci.}\ }\textbf {\bibinfo {volume} {6}},\ \bibinfo {pages} {15} (\bibinfo {year} {1996}{\natexlab{a}})}\BibitemShut {NoStop}%
\bibitem [{\citenamefont {Kresse}\ and\ \citenamefont {Furthm{\"u}ller}(1996{\natexlab{b}})}]{Kresse:1996vf}%
  \BibitemOpen
  \bibfield  {author} {\bibinfo {author} {\bibfnamefont {G.}~\bibnamefont {Kresse}}\ and\ \bibinfo {author} {\bibfnamefont {J.}~\bibnamefont {Furthm{\"u}ller}},\ }\href {\doibase 10.1103/physrevb.54.11169} {\bibfield  {journal} {\bibinfo  {journal} {Phys. Rev. B}\ }\textbf {\bibinfo {volume} {54}},\ \bibinfo {pages} {11169} (\bibinfo {year} {1996}{\natexlab{b}})}\BibitemShut {NoStop}%
\bibitem [{\citenamefont {Perdew}\ \emph {et~al.}(2008)\citenamefont {Perdew}, \citenamefont {Ruzsinszky}, \citenamefont {Csonka}, \citenamefont {Vydrov}, \citenamefont {Scuseria}, \citenamefont {Constantin}, \citenamefont {Zhou},\ and\ \citenamefont {Burke}}]{perdew2008restoring}%
  \BibitemOpen
  \bibfield  {author} {\bibinfo {author} {\bibfnamefont {J.~P.}\ \bibnamefont {Perdew}}, \bibinfo {author} {\bibfnamefont {A.}~\bibnamefont {Ruzsinszky}}, \bibinfo {author} {\bibfnamefont {G.~I.}\ \bibnamefont {Csonka}}, \bibinfo {author} {\bibfnamefont {O.~A.}\ \bibnamefont {Vydrov}}, \bibinfo {author} {\bibfnamefont {G.~E.}\ \bibnamefont {Scuseria}}, \bibinfo {author} {\bibfnamefont {L.~A.}\ \bibnamefont {Constantin}}, \bibinfo {author} {\bibfnamefont {X.}~\bibnamefont {Zhou}}, \ and\ \bibinfo {author} {\bibfnamefont {K.}~\bibnamefont {Burke}},\ }\href {\doibase 10.1103/PhysRevLett.100.136406} {\bibfield  {journal} {\bibinfo  {journal} {Phys. Rev. Lett.}\ }\textbf {\bibinfo {volume} {100}},\ \bibinfo {pages} {136406} (\bibinfo {year} {2008})}\BibitemShut {NoStop}%
\bibitem [{\citenamefont {Anisimov}\ \emph {et~al.}(1991)\citenamefont {Anisimov}, \citenamefont {Zaanen},\ and\ \citenamefont {Andersen}}]{anisimov1991band}%
  \BibitemOpen
  \bibfield  {author} {\bibinfo {author} {\bibfnamefont {V.~I.}\ \bibnamefont {Anisimov}}, \bibinfo {author} {\bibfnamefont {J.}~\bibnamefont {Zaanen}}, \ and\ \bibinfo {author} {\bibfnamefont {O.~K.}\ \bibnamefont {Andersen}},\ }\href {\doibase 10.1103/PhysRevB.44.943} {\bibfield  {journal} {\bibinfo  {journal} {Phys. Rev. B}\ }\textbf {\bibinfo {volume} {44}},\ \bibinfo {pages} {943} (\bibinfo {year} {1991})}\BibitemShut {NoStop}%
\bibitem [{\citenamefont {Dudarev}\ \emph {et~al.}(1998)\citenamefont {Dudarev}, \citenamefont {Botton}, \citenamefont {Savrasov}, \citenamefont {Humphreys},\ and\ \citenamefont {Sutton}}]{Dudarev1998}%
  \BibitemOpen
  \bibfield  {author} {\bibinfo {author} {\bibfnamefont {S.~L.}\ \bibnamefont {Dudarev}}, \bibinfo {author} {\bibfnamefont {G.~A.}\ \bibnamefont {Botton}}, \bibinfo {author} {\bibfnamefont {S.~Y.}\ \bibnamefont {Savrasov}}, \bibinfo {author} {\bibfnamefont {C.~J.}\ \bibnamefont {Humphreys}}, \ and\ \bibinfo {author} {\bibfnamefont {A.~P.}\ \bibnamefont {Sutton}},\ }\href {\doibase 10.1103/PhysRevB.57.1505} {\bibfield  {journal} {\bibinfo  {journal} {Phys. Rev. B}\ }\textbf {\bibinfo {volume} {57}},\ \bibinfo {pages} {1505} (\bibinfo {year} {1998})}\BibitemShut {NoStop}%
\bibitem [{\citenamefont {Bl{\"o}chl}(1994)}]{blochl1994projector}%
  \BibitemOpen
  \bibfield  {author} {\bibinfo {author} {\bibfnamefont {P.~E.}\ \bibnamefont {Bl{\"o}chl}},\ }\href {\doibase 10.1103/PhysRevB.50.17953} {\bibfield  {journal} {\bibinfo  {journal} {Phys. Rev. B}\ }\textbf {\bibinfo {volume} {50}},\ \bibinfo {pages} {17953} (\bibinfo {year} {1994})}\BibitemShut {NoStop}%
\bibitem [{\citenamefont {Kresse}\ and\ \citenamefont {Joubert}(1999)}]{kresse1999ultrasoft}%
  \BibitemOpen
  \bibfield  {author} {\bibinfo {author} {\bibfnamefont {G.}~\bibnamefont {Kresse}}\ and\ \bibinfo {author} {\bibfnamefont {D.}~\bibnamefont {Joubert}},\ }\href {\doibase 10.1103/PhysRevB.59.1758} {\bibfield  {journal} {\bibinfo  {journal} {Phys. Rev. B}\ }\textbf {\bibinfo {volume} {59}},\ \bibinfo {pages} {1758} (\bibinfo {year} {1999})}\BibitemShut {NoStop}%
\bibitem [{\citenamefont {Monkhorst}\ and\ \citenamefont {Pack}(1976)}]{monkhorst1976}%
  \BibitemOpen
  \bibfield  {author} {\bibinfo {author} {\bibfnamefont {H.~J.}\ \bibnamefont {Monkhorst}}\ and\ \bibinfo {author} {\bibfnamefont {J.~D.}\ \bibnamefont {Pack}},\ }\href {\doibase 10.1103/PhysRevB.13.5188} {\bibfield  {journal} {\bibinfo  {journal} {Phys. Rev. B}\ }\textbf {\bibinfo {volume} {13}},\ \bibinfo {pages} {5188} (\bibinfo {year} {1976})}\BibitemShut {NoStop}%
\bibitem [{\citenamefont {Gajdo\v{s}}\ \emph {et~al.}(2006)\citenamefont {Gajdo\v{s}}, \citenamefont {Hummer}, \citenamefont {Kresse}, \citenamefont {Furthm{\"u}ller},\ and\ \citenamefont {Bechstedt}}]{gajdos2006linear}%
  \BibitemOpen
  \bibfield  {author} {\bibinfo {author} {\bibfnamefont {M.}~\bibnamefont {Gajdo\v{s}}}, \bibinfo {author} {\bibfnamefont {K.}~\bibnamefont {Hummer}}, \bibinfo {author} {\bibfnamefont {G.}~\bibnamefont {Kresse}}, \bibinfo {author} {\bibfnamefont {J.}~\bibnamefont {Furthm{\"u}ller}}, \ and\ \bibinfo {author} {\bibfnamefont {F.}~\bibnamefont {Bechstedt}},\ }\href {\doibase 10.1103/PhysRevB.73.045112} {\bibfield  {journal} {\bibinfo  {journal} {Phys. Rev. B}\ }\textbf {\bibinfo {volume} {73}},\ \bibinfo {pages} {045112} (\bibinfo {year} {2006})}\BibitemShut {NoStop}%
\bibitem [{\citenamefont {Baroni}\ and\ \citenamefont {Resta}(1986)}]{baroni1986dielectric}%
  \BibitemOpen
  \bibfield  {author} {\bibinfo {author} {\bibfnamefont {S.}~\bibnamefont {Baroni}}\ and\ \bibinfo {author} {\bibfnamefont {R.}~\bibnamefont {Resta}},\ }\href {\doibase 10.1103/PhysRevB.33.7017} {\bibfield  {journal} {\bibinfo  {journal} {Phys. Rev. B}\ }\textbf {\bibinfo {volume} {33}},\ \bibinfo {pages} {7017} (\bibinfo {year} {1986})}\BibitemShut {NoStop}%
\bibitem [{\citenamefont {Spaldin}(2012)}]{spaldin2012beginner}%
  \BibitemOpen
  \bibfield  {author} {\bibinfo {author} {\bibfnamefont {N.~A.}\ \bibnamefont {Spaldin}},\ }\href {\doibase 10.1016/j.jssc.2012.05.010} {\bibfield  {journal} {\bibinfo  {journal} {J. Solid State Chem.}\ }\textbf {\bibinfo {volume} {195}},\ \bibinfo {pages} {2} (\bibinfo {year} {2012})}\BibitemShut {NoStop}%
\bibitem [{\citenamefont {Henkelman}\ \emph {et~al.}(2000)\citenamefont {Henkelman}, \citenamefont {Uberuaga},\ and\ \citenamefont {J{\'o}nsson}}]{henkelman2000climbing}%
  \BibitemOpen
  \bibfield  {author} {\bibinfo {author} {\bibfnamefont {G.}~\bibnamefont {Henkelman}}, \bibinfo {author} {\bibfnamefont {B.~P.}\ \bibnamefont {Uberuaga}}, \ and\ \bibinfo {author} {\bibfnamefont {H.}~\bibnamefont {J{\'o}nsson}},\ }\href {\doibase 10.1063/1.1329672} {\bibfield  {journal} {\bibinfo  {journal} {J. Chem. Phys.}\ }\textbf {\bibinfo {volume} {113}},\ \bibinfo {pages} {9901} (\bibinfo {year} {2000})}\BibitemShut {NoStop}%
\bibitem [{\citenamefont {Zhang}\ and\ \citenamefont {Northrup}(1991)}]{zhang1991chemical}%
  \BibitemOpen
  \bibfield  {author} {\bibinfo {author} {\bibfnamefont {S.~B.}\ \bibnamefont {Zhang}}\ and\ \bibinfo {author} {\bibfnamefont {J.~E.}\ \bibnamefont {Northrup}},\ }\href {\doibase 10.1103/PhysRevLett.67.2339} {\bibfield  {journal} {\bibinfo  {journal} {Phys. Rev. Lett.}\ }\textbf {\bibinfo {volume} {67}},\ \bibinfo {pages} {2339} (\bibinfo {year} {1991})}\BibitemShut {NoStop}%
\bibitem [{\citenamefont {Freysoldt}\ \emph {et~al.}(2014)\citenamefont {Freysoldt}, \citenamefont {Grabowski}, \citenamefont {Hickel}, \citenamefont {Neugebauer}, \citenamefont {Kresse}, \citenamefont {Janotti},\ and\ \citenamefont {Van~de Walle}}]{freysoldt2014first}%
  \BibitemOpen
  \bibfield  {author} {\bibinfo {author} {\bibfnamefont {C.}~\bibnamefont {Freysoldt}}, \bibinfo {author} {\bibfnamefont {B.}~\bibnamefont {Grabowski}}, \bibinfo {author} {\bibfnamefont {T.}~\bibnamefont {Hickel}}, \bibinfo {author} {\bibfnamefont {J.}~\bibnamefont {Neugebauer}}, \bibinfo {author} {\bibfnamefont {G.}~\bibnamefont {Kresse}}, \bibinfo {author} {\bibfnamefont {A.}~\bibnamefont {Janotti}}, \ and\ \bibinfo {author} {\bibfnamefont {C.~G.}\ \bibnamefont {Van~de Walle}},\ }\href {\doibase 10.1103/RevModPhys.86.253} {\bibfield  {journal} {\bibinfo  {journal} {Rev. Mod. Phys.}\ }\textbf {\bibinfo {volume} {86}},\ \bibinfo {pages} {253} (\bibinfo {year} {2014})}\BibitemShut {NoStop}%
\bibitem [{\citenamefont {Lany}\ and\ \citenamefont {Zunger}(2009)}]{lany2009accurate}%
  \BibitemOpen
  \bibfield  {author} {\bibinfo {author} {\bibfnamefont {S.}~\bibnamefont {Lany}}\ and\ \bibinfo {author} {\bibfnamefont {A.}~\bibnamefont {Zunger}},\ }\href {\doibase 10.1088/0965-0393/17/8/084002} {\bibfield  {journal} {\bibinfo  {journal} {Model. Simul. Mater. Sci. Eng.}\ }\textbf {\bibinfo {volume} {17}},\ \bibinfo {pages} {084002} (\bibinfo {year} {2009})}\BibitemShut {NoStop}%
\bibitem [{\citenamefont {Ong}\ \emph {et~al.}(2013)\citenamefont {Ong}, \citenamefont {Richards}, \citenamefont {Jain}, \citenamefont {Hautier}, \citenamefont {Kocher}, \citenamefont {Cholia}, \citenamefont {Gunter}, \citenamefont {Chevrier}, \citenamefont {Persson},\ and\ \citenamefont {Ceder}}]{ong2013pymatgen}%
  \BibitemOpen
  \bibfield  {author} {\bibinfo {author} {\bibfnamefont {S.~P.}\ \bibnamefont {Ong}}, \bibinfo {author} {\bibfnamefont {W.~D.}\ \bibnamefont {Richards}}, \bibinfo {author} {\bibfnamefont {A.}~\bibnamefont {Jain}}, \bibinfo {author} {\bibfnamefont {G.}~\bibnamefont {Hautier}}, \bibinfo {author} {\bibfnamefont {M.}~\bibnamefont {Kocher}}, \bibinfo {author} {\bibfnamefont {S.}~\bibnamefont {Cholia}}, \bibinfo {author} {\bibfnamefont {D.}~\bibnamefont {Gunter}}, \bibinfo {author} {\bibfnamefont {V.~L.}\ \bibnamefont {Chevrier}}, \bibinfo {author} {\bibfnamefont {K.~A.}\ \bibnamefont {Persson}}, \ and\ \bibinfo {author} {\bibfnamefont {G.}~\bibnamefont {Ceder}},\ }\href {\doibase 10.1016/j.commatsci.2012.10.028} {\bibfield  {journal} {\bibinfo  {journal} {Comput. Mater. Sci.}\ }\textbf {\bibinfo {volume} {68}},\ \bibinfo {pages} {314} (\bibinfo {year} {2013})}\BibitemShut {NoStop}%
\bibitem [{\citenamefont {Morgan}()}]{pyscfermi}%
  \BibitemOpen
  \bibfield  {author} {\bibinfo {author} {\bibfnamefont {B.}~\bibnamefont {Morgan}},\ }\href {https://github.com/bjmorgan/py-sc-fermi} {\enquote {\bibinfo {title} {py-sc-fermi},}\ }\BibitemShut {NoStop}%
\bibitem [{\citenamefont {Ichibha}\ \emph {et~al.}(2023)\citenamefont {Ichibha}, \citenamefont {Saritas}, \citenamefont {Krogel}, \citenamefont {Luo}, \citenamefont {Kent},\ and\ \citenamefont {Reboredo}}]{ichibha2023existence}%
  \BibitemOpen
  \bibfield  {author} {\bibinfo {author} {\bibfnamefont {T.}~\bibnamefont {Ichibha}}, \bibinfo {author} {\bibfnamefont {K.}~\bibnamefont {Saritas}}, \bibinfo {author} {\bibfnamefont {J.~T.}\ \bibnamefont {Krogel}}, \bibinfo {author} {\bibfnamefont {Y.}~\bibnamefont {Luo}}, \bibinfo {author} {\bibfnamefont {P.~R.}\ \bibnamefont {Kent}}, \ and\ \bibinfo {author} {\bibfnamefont {F.~A.}\ \bibnamefont {Reboredo}},\ }\href {\doibase 10.1038/s41598-023-33578-1} {\bibfield  {journal} {\bibinfo  {journal} {Sci. Rep.}\ }\textbf {\bibinfo {volume} {13}},\ \bibinfo {pages} {6703} (\bibinfo {year} {2023})}\BibitemShut {NoStop}%
\bibitem [{\citenamefont {Shannon}(1976)}]{shannon1976revised}%
  \BibitemOpen
  \bibfield  {author} {\bibinfo {author} {\bibfnamefont {R.~D.}\ \bibnamefont {Shannon}},\ }\href {\doibase 10.1107/S0567739476001551} {\bibfield  {journal} {\bibinfo  {journal} {Acta Cryst. A}\ }\textbf {\bibinfo {volume} {32}},\ \bibinfo {pages} {751} (\bibinfo {year} {1976})}\BibitemShut {NoStop}%
\bibitem [{\citenamefont {Goodenough}(1955)}]{goodenough1955theory}%
  \BibitemOpen
  \bibfield  {author} {\bibinfo {author} {\bibfnamefont {J.~B.}\ \bibnamefont {Goodenough}},\ }\href {\doibase 10.1103/PhysRev.100.564} {\bibfield  {journal} {\bibinfo  {journal} {Phys. Rev.}\ }\textbf {\bibinfo {volume} {100}},\ \bibinfo {pages} {564} (\bibinfo {year} {1955})}\BibitemShut {NoStop}%
\bibitem [{\citenamefont {Goodenough}(1958)}]{goodenough1958interpretation}%
  \BibitemOpen
  \bibfield  {author} {\bibinfo {author} {\bibfnamefont {J.~B.}\ \bibnamefont {Goodenough}},\ }\href {\doibase 10.1016/0022-3697(58)90107-0} {\bibfield  {journal} {\bibinfo  {journal} {Journal of Physics and Chemistry of Solids}\ }\textbf {\bibinfo {volume} {6}},\ \bibinfo {pages} {287} (\bibinfo {year} {1958})}\BibitemShut {NoStop}%
\bibitem [{\citenamefont {Kanamori}(1959)}]{kanamori1959superexchange}%
  \BibitemOpen
  \bibfield  {author} {\bibinfo {author} {\bibfnamefont {J.}~\bibnamefont {Kanamori}},\ }\href {\doibase 10.1016/0022-3697(59)90061-7} {\bibfield  {journal} {\bibinfo  {journal} {Journal of Physics and Chemistry of Solids}\ }\textbf {\bibinfo {volume} {10}},\ \bibinfo {pages} {87} (\bibinfo {year} {1959})}\BibitemShut {NoStop}%
\bibitem [{\citenamefont {Rabe}\ \emph {et~al.}(2007)\citenamefont {Rabe}, \citenamefont {Dawber}, \citenamefont {Lichtensteiger}, \citenamefont {Ahn},\ and\ \citenamefont {Triscone}}]{rabe2007modern}%
  \BibitemOpen
  \bibfield  {author} {\bibinfo {author} {\bibfnamefont {K.~M.}\ \bibnamefont {Rabe}}, \bibinfo {author} {\bibfnamefont {M.}~\bibnamefont {Dawber}}, \bibinfo {author} {\bibfnamefont {C.}~\bibnamefont {Lichtensteiger}}, \bibinfo {author} {\bibfnamefont {C.~H.}\ \bibnamefont {Ahn}}, \ and\ \bibinfo {author} {\bibfnamefont {J.-M.}\ \bibnamefont {Triscone}},\ }\enquote {\bibinfo {title} {Modern physics of ferroelectrics: {E}ssential background},}\ in\ \href {\doibase 10.1007/978-3-540-34591-6_1} {\emph {\bibinfo {booktitle} {Physics of Ferroelectrics: A Modern Perspective}}}\ (\bibinfo  {publisher} {Springer},\ \bibinfo {address} {Berlin, Heidelberg},\ \bibinfo {year} {2007})\ pp.\ \bibinfo {pages} {1--30}\BibitemShut {NoStop}%
\bibitem [{\citenamefont {Beckman}\ \emph {et~al.}(2009)\citenamefont {Beckman}, \citenamefont {Wang}, \citenamefont {Rabe},\ and\ \citenamefont {Vanderbilt}}]{beckman2009ideal}%
  \BibitemOpen
  \bibfield  {author} {\bibinfo {author} {\bibfnamefont {S.}~\bibnamefont {Beckman}}, \bibinfo {author} {\bibfnamefont {X.}~\bibnamefont {Wang}}, \bibinfo {author} {\bibfnamefont {K.~M.}\ \bibnamefont {Rabe}}, \ and\ \bibinfo {author} {\bibfnamefont {D.}~\bibnamefont {Vanderbilt}},\ }\href {\doibase 10.1103/PhysRevB.79.144124} {\bibfield  {journal} {\bibinfo  {journal} {Phys. Rev. B}\ }\textbf {\bibinfo {volume} {79}},\ \bibinfo {pages} {144124} (\bibinfo {year} {2009})}\BibitemShut {NoStop}%
\bibitem [{\citenamefont {Chen}\ \emph {et~al.}(2025)\citenamefont {Chen}, \citenamefont {Jiang}, \citenamefont {Zhang}, \citenamefont {Qiu}, \citenamefont {Liang}, \citenamefont {Zhang}, \citenamefont {Zhu}, \citenamefont {Ohkochi}, \citenamefont {Chen}, \citenamefont {Wang}, \citenamefont {Liu}, \citenamefont {He}, \citenamefont {Ma}, \citenamefont {Yu}, \citenamefont {Lin}, \citenamefont {Nan},\ and\ \citenamefont {Yi}}]{chen2025pure}%
  \BibitemOpen
  \bibfield  {author} {\bibinfo {author} {\bibfnamefont {H.}~\bibnamefont {Chen}}, \bibinfo {author} {\bibfnamefont {D.}~\bibnamefont {Jiang}}, \bibinfo {author} {\bibfnamefont {Y.}~\bibnamefont {Zhang}}, \bibinfo {author} {\bibfnamefont {X.}~\bibnamefont {Qiu}}, \bibinfo {author} {\bibfnamefont {Y.}~\bibnamefont {Liang}}, \bibinfo {author} {\bibfnamefont {Q.}~\bibnamefont {Zhang}}, \bibinfo {author} {\bibfnamefont {F.}~\bibnamefont {Zhu}}, \bibinfo {author} {\bibfnamefont {T.}~\bibnamefont {Ohkochi}}, \bibinfo {author} {\bibfnamefont {M.}~\bibnamefont {Chen}}, \bibinfo {author} {\bibfnamefont {Y.}~\bibnamefont {Wang}}, \bibinfo {author} {\bibfnamefont {J.}~\bibnamefont {Liu}}, \bibinfo {author} {\bibfnamefont {Q.}~\bibnamefont {He}}, \bibinfo {author} {\bibfnamefont {J.}~\bibnamefont {Ma}}, \bibinfo {author} {\bibfnamefont {P.}~\bibnamefont {Yu}}, \bibinfo {author} {\bibfnamefont {Y.}~\bibnamefont {Lin}}, \bibinfo {author} {\bibfnamefont {T.}~\bibnamefont {Nan}}, \ and\ \bibinfo {author} {\bibfnamefont {D.}~\bibnamefont {Yi}},\ }\href {\doibase 10.1038/s41467-025-61490-x} {\bibfield  {journal} {\bibinfo  {journal} {Nat. Commun.}\ }\textbf {\bibinfo {volume} {16}},\ \bibinfo {pages} {6257} (\bibinfo {year} {2025})}\BibitemShut {NoStop}%
\end{thebibliography}%


\begin{thebibliography}{1}%
\makeatletter
\providecommand \@ifxundefined [1]{%
 \@ifx{#1\undefined}
}%
\providecommand \@ifnum [1]{%
 \ifnum #1\expandafter \@firstoftwo
 \else \expandafter \@secondoftwo
 \fi
}%
\providecommand \@ifx [1]{%
 \ifx #1\expandafter \@firstoftwo
 \else \expandafter \@secondoftwo
 \fi
}%
\providecommand \natexlab [1]{#1}%
\providecommand \enquote  [1]{``#1''}%
\providecommand \bibnamefont  [1]{#1}%
\providecommand \bibfnamefont [1]{#1}%
\providecommand \citenamefont [1]{#1}%
\providecommand \href@noop [0]{\@secondoftwo}%
\providecommand \href [0]{\begingroup \@sanitize@url \@href}%
\providecommand \@href[1]{\@@startlink{#1}\@@href}%
\providecommand \@@href[1]{\endgroup#1\@@endlink}%
\providecommand \@sanitize@url [0]{\catcode `\\12\catcode `\$12\catcode `\&12\catcode `\#12\catcode `\^12\catcode `\_12\catcode `\%12\relax}%
\providecommand \@@startlink[1]{}%
\providecommand \@@endlink[0]{}%
\providecommand \url  [0]{\begingroup\@sanitize@url \@url }%
\providecommand \@url [1]{\endgroup\@href {#1}{\urlprefix }}%
\providecommand \urlprefix  [0]{URL }%
\providecommand \Eprint [0]{\href }%
\providecommand \doibase [0]{http://dx.doi.org/}%
\providecommand \selectlanguage [0]{\@gobble}%
\providecommand \bibinfo  [0]{\@secondoftwo}%
\providecommand \bibfield  [0]{\@secondoftwo}%
\providecommand \translation [1]{[#1]}%
\providecommand \BibitemOpen [0]{}%
\providecommand \bibitemStop [0]{}%
\providecommand \bibitemNoStop [0]{.\EOS\space}%
\providecommand \EOS [0]{\spacefactor3000\relax}%
\providecommand \BibitemShut  [1]{\csname bibitem#1\endcsname}%
\let\auto@bib@innerbib\@empty
%</preamble>
\bibitem [{\citenamefont {Xiang}\ \emph {et~al.}(2012)\citenamefont {Xiang}, \citenamefont {Lee}, \citenamefont {Koo}, \citenamefont {Gong},\ and\ \citenamefont {Whangbo}}]{xiang2012magnetic}%
  \BibitemOpen
  \bibfield  {author} {\bibinfo {author} {\bibfnamefont {H.}~\bibnamefont {Xiang}}, \bibinfo {author} {\bibfnamefont {C.}~\bibnamefont {Lee}}, \bibinfo {author} {\bibfnamefont {H.-J.}\ \bibnamefont {Koo}}, \bibinfo {author} {\bibfnamefont {X.}~\bibnamefont {Gong}}, \ and\ \bibinfo {author} {\bibfnamefont {M.-H.}\ \bibnamefont {Whangbo}},\ }\href {\doibase 10.1039/C2DT31662E} {\bibfield  {journal} {\bibinfo  {journal} {Dalton Trans.}\ }\textbf {\bibinfo {volume} {42}},\ \bibinfo {pages} {823} (\bibinfo {year} {2012})}\BibitemShut {NoStop}%
\end{thebibliography}%
%%%%%%%%%%%%%%%%%%%%%%%%%%%%%%%%%%%%%%%%%%%%%%%%%%%%%%%%%%%%%%%%%%%%%%%%%%%%%%%%%

\clearpage
\clearpage
\setcounter{page}{1}
\renewcommand{\thetable}{S\arabic{table}} 
\setcounter{table}{0}
\renewcommand{\thefigure}{S\arabic{figure}}
\setcounter{figure}{0}
\renewcommand{\thesection}{S\arabic{section}}
\setcounter{section}{0}
\renewcommand{\theequation}{S\arabic{equation}}
\setcounter{equation}{0}
\onecolumngrid

\begin{center}
\textbf{Supplementary information for\\\vspace{0.5 cm}
\large Multiferroicity and 180$^\circ$ domain switching in \ce{LaFeO3} via Antisite Defects\\\vspace{0.3 cm}}
Souren Majani and Ulrich Aschauer

\small
\textit{Department of Chemistry and Physics of Materials, University of Salzburg, Jakob-Haringer-Strasse 2a, 5020 Salzburg, Austria}

(Dated: \today)
\end{center}

\section{Magnetic exchange constants}

We have calculated the Heisenberg coupling constant to confirm the structural Kanamori-Goodenough picture in the main text by the solving the following Hamiltonian~\citeSI{xiang2012magnetic} for all Fe--O--Fe spin dimers:
\begin{equation} 	
\hat{H}_{\text{spin}} = J \hat{\mathbf{S}}_1 \cdot \hat{\mathbf{S}}_2, \tag{3}
\end{equation}
where $J$ is the Heisenberg spin exchange constant and $\hat{S}_1$ and $\hat{S}_2$ represent the spin operators of the \ce{Fe_{La}} and its first nearest neighbors. From the $J$ values in Table~\ref{tab:Exch_Angles} it can be seen that all the exchange interactions are positive (i.e. antiferromagnatic) showing a strong preference for an antiparallel alignment of the antisite spin with all surrounding lattice Fe atoms. We can, however see that Fe3, for which the Fe--O--Fe angle with the antisite is largest, also has the largest $J$. As such, this exchange will dominate the spin direction of the antisite as discussed in the main text.

\begin{table}[h]
    \centering
    \caption{Heisenberg exchange constants and Fe--O--Fe angles of the \ce{Fe_{La}} in configurations 3 and 4 with its nearest neighbor lattice Fe atoms.}
    \label{tab:Exch_Angles}
    \begin{tabular}{llccc}
        \hline
        Config. 3 & Neighbor & J (eV) & Angle ($^\circ$) \\
        \hline
        & Fe3  & 0.03559 & 125.46      \\
        & Fe4  & 0.02282 & 118.68      \\
        & Fe9  & 0.01900 & 113.28      \\
        & Fe6  & 0.01239 & 107.03      \\
        & Fe8  & 0.00818 & 87.29-87.62 \\
        & Fe7  & 0.00379 & 90.83-91.06 \\
        & Fe5  & 0.00221 &  -          \\
        & Fe2  & 0.00183 &  -          \\
        \hline
        Config. 4 & Neighbor & J (eV) & Angle ($^\circ$) \\
        \hline
        & Fe7  & 0.03554 & 125.47      \\
        & Fe8  & 0.02270 & 118.60      \\
        & Fe5  & 0.01905 & 113.34      \\
        & Fe2  & 0.01234 & 106.97      \\
        & Fe4  & 0.00813 & 87.55/87.25 \\
        & Fe3  & 0.00379 & 91.07/90.87 \\
        & Fe9  & 0.00221 &  -          \\
        & Fe6  & 0.00182 &  -          \\
        \hline   
        \end{tabular}
\end{table}

\newpage
\section{Switching barriers and polarization}

\begin{table}[h]
    \centering
    \caption{Distance of the antisite from the centrosymmetric position, polarization due to antisite displacement and the total polarization.}
    \label{tab:pol}
    \begin{tabular}{lccc}
        \hline
        Config. & $d$ (\AA) & $P_{\ce{Fe_{La}}}$ ($\mu$C/cm$^2$) & $P$ ($\mu$C/cm$^2$) \\
        \hline
        1 & 1.048 & 11.204 & 17.901 \\
        2 & 0.198 & 2.245  & 4.646 \\
        3 & 0.757 & 8.195  & 16.155 \\
        4 & 0.755 & 8.192  & 13.688 \\
        \hline
    \end{tabular}
\end{table}

\begin{figure}[h]
  \includegraphics{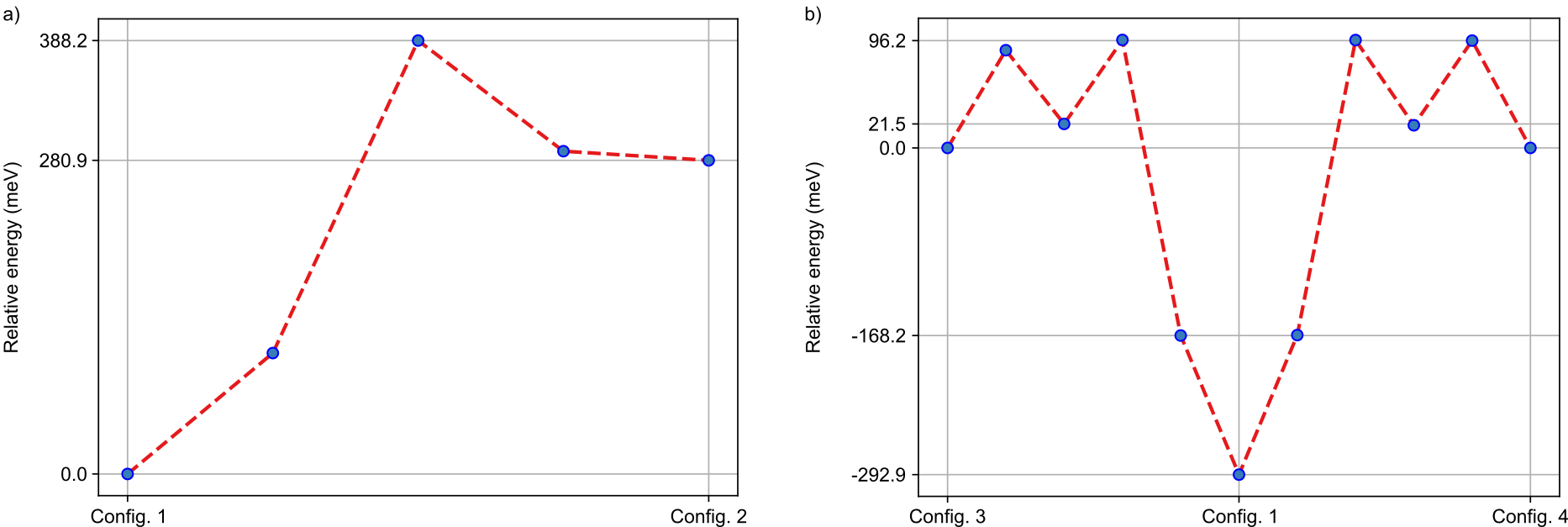}
  \caption{NEB energy profiles for a) switching from configuration 1 to 2 and b) switching between configurations 1, 3 and 4.}
\end{figure}

\begin{table}[h]
    \centering
    \caption{Energy barriers for polarization switching obtained from NEB calculations and coercive fields required for polarization switching. For the latter, we have used the component of the polarization primarily directed along the switching direction, i.e. $z$ for switching from/to configuration 1 and $y$ for all others.}
    \label{tab:NEB}
    \begin{tabular}{lcc}
        \hline
        Transition path & $E_b$ (meV) & $\epsilon_{c}$ (MV/m) \\
        \hline
        ${1\rightarrow 2}$   & 388 & 1115 \\
        ${2\rightarrow 3,4}$ & 136 & 1506 \\
        ${3\rightarrow 2}$   & 124 & 395  \\
        ${4\rightarrow 2}$   & 124 & 466  \\
        ${1\rightarrow 3,4}$ & 389 & 1275 \\
        ${3\rightarrow 1}$   & 96  & 348  \\
        ${4\rightarrow 1}$   & 96  & 437  \\
		${2\rightarrow 1}$   & 107 & 1911 \\
        \hline
    \end{tabular}
\end{table}

\bibliographystyleSI{apsrev4-1}
\bibliographySI{references}

\end{document}